\documentclass[twocolumn,showpacs,amsmath,amssymb]{revtex4}
\usepackage{graphicx}

\begin{document}
\title{Janus particles with coupled electric and magnetic moments make a disordered magneto-electric medium}
\author{Ambarish Ghosh}
\affiliation{The Rowland Institute at Harvard, Harvard University,
Cambridge, Massachusetts 02142}
\author{Nicholas K. Sheridon$^\ddagger$}
\affiliation{The Rowland Institute at Harvard, Harvard University,
Cambridge, Massachusetts 02142}
\author{Peer Fischer}
\affiliation{The Rowland Institute at Harvard, Harvard University,
Cambridge, Massachusetts 02142}

\begin{abstract}
We demonstrate that by combining permanent electric and magnetic
moments in particles, it is possible to realize a new type of
medium that allows for a cross-correlation between electric and
magnetic properties of matter, known as magnetoelectric coupling.
Magnetoelectric materials have so far been restricted to systems
that exhibit long-range order in their electric and magnetic
moments. Here, we show that a room-temperature, switchable
magnetoelectric can be realized that is naturally disordered. The
building blocks are Tellegen particles that orient in either an
electric or a magnetic field.
\end{abstract}
 \maketitle

For a static electric field to give rise to a magnetization and
similarly for a static magnetic field to induce an electric
polarization in matter requires materials that exhibit special
symmetries. Known magnetoelectric materials are single-phase
non-centrosymmetric magnetic crystals or composites that contain a
piezoelectric phase \cite{fiebig2005,eerenstein2006,nan1994}. Most
materials, however, are naturally disordered and so it is
important to establish whether a disordered medium can support
magnetoelectric (ME) phenomena. Conceivably the most general,
random system is a gas or a liquid. If it consists of electric
dipolar molecules, then application of an electric field would
exert a torque and align the molecules with the field. Should the
molecules also possess a magnetic moment parallel to the axis of
the electric dipole moment, then orienting one moment should fix
the other in space, provided the electric and magnetic dipoles are
`tied together'. However, van Vleck showed that this simple idea
cannot be realized with molecules, such as nitric oxide (NO), even
though NO is electric dipolar and has a magnetic moment
\cite{Vleck1985}. The reason is that the magnetic moment in
paramagnetic molecules is not of fixed orientation in the absence
of a magnetic field \cite{Vleck1985,tn1}. It follows that no
magnetization can be induced in such an isotropic
molecular medium, when only an electric field is applied. \\
The idea of coupled electric and magnetic dipole moments that
preferentially align with a field may, however, still be realized,
if the ME building-block is an electric-dipolar ferromagnetic
particle as seen in Figure 1a. Once magnetized, each particle
carries its own magnetic field which renders it anti-symmetric
under time reversal symmetry and thus allows for ME effects
\cite{tn1}. An isotropic medium made of particles that have
coupled permanent electric and magnetic moments (see Figure 1b)
has first been considered by Tellegen, when he conceived of a
fifth circuit element of an electrical network, the ``gyrator"
\cite{tellegen1948,odell}. Remarkably, Tellegen's proposal to make
orientable microscopic magnets coupled to electrets has never been
realized before \cite{fiebig2005,raab1997}. A few experimental
studies consider Tellegen's proposal, but none of these are based
on particles with permanent coupled electric and magnetic dipole
moments \cite{saha2002,tretyakov2003,zhai2006}.
\begin{figure}
\includegraphics{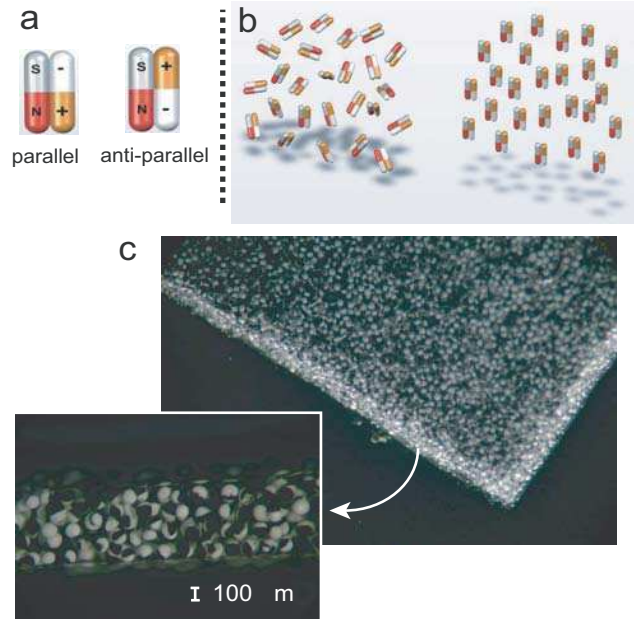}
\caption{\label{fig1} (color online)
 Magnetoelectic particles:
(a) Schematic of a magnetoelectric particle with electric and
dipole moments parallel (left) and anti-parallel (right)
\cite{lindell1994}. (b) Magetoelectric medium with anti-parallel
moments in a disordered state (left) and when it is ordered by
application of
 either an external electric or a magnetic field (right). (c) Image of an
elastomer sheet containing ME bichromal particles. The inset shows
a magnified cross section of the sheet.}
\end{figure}
Here we show that magnetoelectric particles can be made and,
moreover, that they can be dispersed randomly in a matrix to make
a disordered material that exhibits a sizeable ME response. In
this new class of materials the ME effect is based on the dynamics
of the microscopic
particles. \\
There are many examples of micro- and nanoparticles that are
either electric or magnetic dipolar. For the present study it is
required that both moments are present and that they are `tied
together'. Further, it is convenient if these particles are also
optically anisotropic, as this can serve as an independent measure
of the particles' alignment with an applied field
\cite{takei1997,anker2003} and hence their ME response. Two-faced,
``Janus" particles \cite{perro2005} that are separately
electrically dipolar
\cite{takei1997,crowley2002,nisisako2006,cayre2003} or
ferromagnetic \cite{anker2003,correa-duarte2005} have been made.
It should thus be possible to combine these characteristics. This
is indeed the case, and we have for instance used fluorescent
silica particles to make magnetoelectric Janus particles
\cite{tn1}. However, our aim here is not only to make
magnetoelectric particles, but to also obtain and measure
appreciable induced bulk magnetizations and electric
polarizations. This requires large number densities and it becomes
necessary to disperse the particles such that they do not interact
too strongly with each other, since large dipole moments in close
proximity favor an anti-parallel coupling and would thus preclude
orientation by an external field. It is therefore advantageous to
disperse the magnetoelectric particles in a suitable medium that
acts as a spacer and prevents agglomeration of the particles,
while still allowing for their rotation. For these reasons the
magnetoelectric medium of this Letter is based on a system used
for electronic paper (Gyricon)
\cite{sheridon2005}. \\
\begin{figure}[b]
\includegraphics{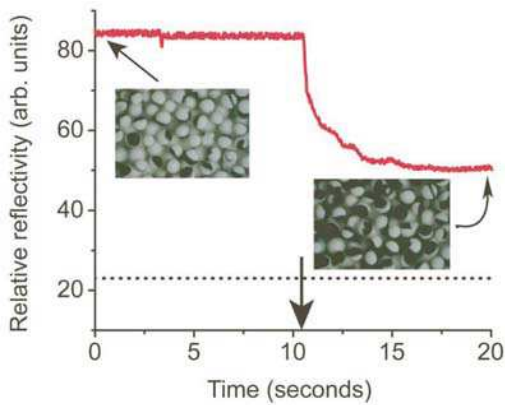}
\caption{\label{fig2} (color online) Typical randomization of the
elastomer sheet containing magnetoelectric beads. A voltage is
applied and the beads are first predominately white-side up (image
on the left). After the electric field is switched off (vertical
marker), the beads quickly reorient to form a disordered medium
(image on the right). The speed and dynamics of the reorientation
depends on the size of the fluid cavity and the strength of the
applied electric field. The dotted line indicates the reflectivity
for black-side up. Similar switching can be observed visually when
a magnetic field is applied.}
\end{figure}
\begin{figure}
\includegraphics{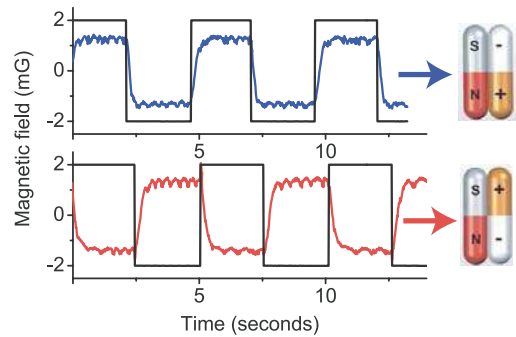}
\caption{\label{fig3} (color online) Measured magnetic field as an
electric field (voltage) is applied across the ME material. A 0.2
Hz square wave of 350 volts (black line, no scale indicated) is
applied to the sheets of Fig. 1. In one sheet the particles'
electric and magnetic moments are parallel (top) and in the other
they are anti-parallel (bottom). The electric field induces a
magnetization which give rises to a magnetic field (where 1mG  =
10$^{-7}$ T) that is in phase (parallel, blue trace) and out of
phase (antiparallel, red trace) with the applied electric field.}
\end{figure}
The system consists of spherical ~100 $\mu$m polyethylene
particles which are isotropically dispersed in a sheet of
elastomer such that each particle occupies a small fluid cavity in
which it is free to rotate \cite{crowley2002,sheridon2005}. Each
bead-like particle is electrically dipolar as the two
differently-colored hemispheres have opposite electric charges.
Here we show that these beads can also be made to carry a
permanent magnetic moment, if made with a ferromagnetic
pigment/powder \cite{tn1}. The strength of the magnetic moment can
be adjusted by varying the amount and nature of the ferromagnetic
dopant. Once dispersed in the elastomer, the beads are oriented by
an electric field and magnetized, such that the permanent electric
and magnetic moments of each bead are either parallel, or
anti-parallel. The sheet now contains randomly
dispersed magnetoelectric particles (Fig. 1c). \\
Because the axis of the optical anisotropy of the beads is also
the axis of their electric and magnetic moments, the measured
intensity of reflected light can serve as an optical measure of
the average orientation of the beads and thus the total ME
response. In the present ME system, the particle interactions are
weak enough to permit orientation by the field, but strong enough
to cause the beads to randomize in the absence of the field. In
Figure 2 it is seen that once the aligning field has been switched
off, the sheet turns from white to gray as the beads become
randomly oriented. In the absence of any field the medium may thus
be characterized as being disordered and isotropic, yet containing
particles that exhibit magnetoelectric coupling. This is in
contrast to all known magnetoelectric media that even in the
absence of any fields contain at least one phase that is ordered
\cite{fiebig2005,eerenstein2006,nan1994,ryu2002}.
In addition to optical measurements, we are able to directly
record the ME response of the medium (further information is
available online \cite{tn1}). The sheet is contained between two
conductive transparent windows that make a capacitor and a voltage
may be applied across the capacitor such that the oriented
particles give rise to a magnetic field. This is detected with a
fluxgate magnetometer. The functional form of the detected
magnetic field closely follows the applied electric field
(voltage), as can be seen in Figure 3. Similarly, upon application
of a magnetic field provided by a solenoid, a voltage is recorded
on an electrometer.
\begin{figure}
\includegraphics{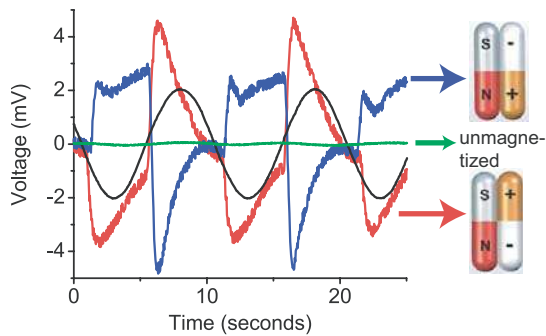}
\caption{\label{fig3} (color online) Voltage resulting from
application of a  sinusoidal magnetic field with an amplitude of
90 Gauss at 0.1 Hz (black line, no scale indicated). The
unmagnetized sheet does not give rise to a voltage (green trace).
The same sheet is now magnetized such that the electric and
magnetic moments of the ME particles are parallel (blue trace),
and then re-magnetizationed such that the moments are
anti-parallel (red trace). The slow driving field ensures that
pick-up is minimal, but at these frequencies an asymmetry in the
bead rotation can be observed, which is explained by the presence
of shallowly trapped charges in the adhesion layer \cite{tn1}.}
\end{figure}
The electrical detection is more challenging, and it is seen that
the detected voltage does not exactly follow the applied magnetic
field, as leakage currents cause a loss of charge and hence
voltage before the magnetic field reverses sign, as is seen in
Figure 4. The unmagnetized sheet exhibits no ME effect, whereas
the parallel and anti-parallel particles give rise to a ME
response that is of approximately
equal magnitude and of opposite sign. \\
To understand the ME response in more detail, one needs to
consider the dynamics of the ME beads. The orientation of the
beads is to first approximation described by equating the torque
due to the applied field with the rotational drag of the particles
\cite{tn1}. In addition, the particles experience adhesion forces
when they come into contact with the walls of the cavity. However,
each bead also possesses a net positive charge. When a large
enough voltage is applied to the medium, then the charge permits
the particles to be pulled away from the cavity wall so that they
are free to rotate. If one considers a collection of identical
particles, then, in the limit of static fields, one would expect
the response of the sheet to be described by a step function: no
magnetization for fields that are too small to overcome the
adhesion force, and otherwise the full saturation magnetization.
Both the off-set voltages as well as the saturation magnetization
(corresponding to complete alignment of the beads) is seen in the
experimental data shown in Figure 5a.
\begin{figure}
\includegraphics{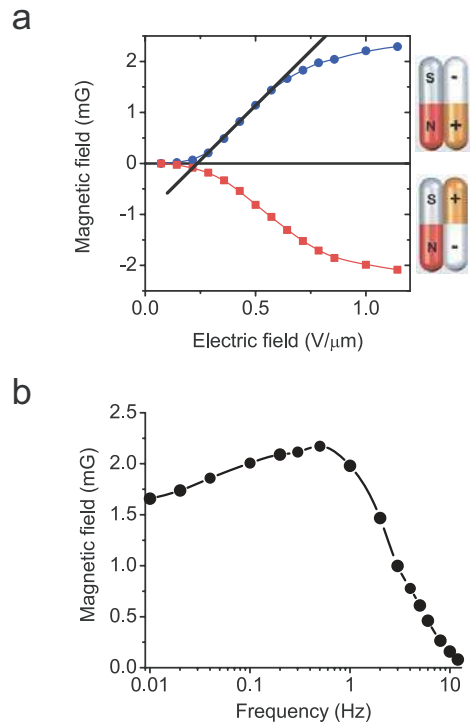}
\caption{\label{fig4} (color online) (a) Measured magnetic field
as a function of a 1 Hz sinusoidal electric field. The detected
magnetizations are in phase (parallel: blue dots) and out of phase
(antiparallel: red squares) with the applied electric fields. The
parallel and antiparallel moments give rise to magnetizations of
opposite sign for the same applied electric field. The straight
line is a linear fit to the approximately linear region of the
graph. (b) Indiced magnetic field as a function of the frequency
of the applied electric field. At high frequency the rotational
drag is larger than the applied resulting in incomplete rotation
of the ME particles and therefore diminished induced
magnetizations.  For lower frequencies there is a slight reduction
in magnetization as here the adhesion of the beads becomes
stronger the longer the beads are in contact with the walls of the
cavity. Note that magnetic fields due to currents and not due to
the ME effect, become larger with increasing frequency, and are
negligible in the present setup. }
\end{figure}
The intermediate region of the graph is sigmoidal, which is partly
explained by the variance of the bead properties and hence the
distribution of adhesion potentials. If we only consider the
region that exhibits approximately linear magnetoelectric
coupling, and write
\begin{equation}
\vec{M} \approx \chi \vec{E}  \;\;\text{and}\;\; \vec{P} \approx
\chi \vec{B} \;,
\end{equation}
where $\vec{M}$ is the bulk magnetization and $\vec{P}$ the
polarization, with $\vec{E}$ and $\vec{B}$, respectively, the
electric and magnetic fields, then we find from a linear fit to
the data that $\chi$=0.4 ps/m (=0.07 mV/(cm Oe) = 3$\times
10^{-7}$ A/V). Surprisingly, the magnetoelectric coefficient of
our isotropic ME medium is already not much smaller than
coefficients found in ME crystals, such as chromium oxide, where
$\chi_{zz}$=4.1 ps/m at ~270 K \cite{fiebig2005}. The magnetic
moment of the ME beads, and hence the ME effect, can be
significantly increased without compromising the functionality of
the material. \\
We note that the medium exhibits dynamical hysteresis, i.e. the
graph in Figure 5a is a function of frequency, as is seen in the
measurements shown in Figure 5b. \\
Tellegen assumed that the
particles with coupled permanent electric and magnetic moments
should behave linearly, as this is advantageous for a network
element \cite{tellegen1948}. It is important to realize that the
ME particles need to interact with each other or their
surroundings as otherwise they could not be re-oriented by the
field once they are all perfectly aligned. The nature of these
interactions will therefore determine the particular functional
form of the ME response. This will in general not be linear, as is
for instance seen in Fig. 5a. One might, nevertheless, expect that
the present system can be operated in its linear range by
application of a dc bias and a small ac field of an appropriate
frequency, as is for instance done in ordered ME composites that
exhibit a nonlinear ME effect \cite{ryu2002}. However, we found
the statistical distribution of the bead release times from the
cavity wall was too large to allow this possibility to be tested
with the current system. Interestingly, the question of whether
the magnetoelectric particles envisioned by Tellegen, can give
rise to a linear ME effect is the subject of an ongoing
theoretical debate \cite{lakhtakia1994b,raab1997,hehl2005}. Now,
that we have shown that ME particles according to Tellegen's
recipe can be used to make a ME medium, it will be a topic of
future research to establish if this or a related system can be
devised which permits
the observation of a linear ME effect. \\
In summray, we have made particles with coupled permanent magnetic
and electric moments and used these to fabricate a switcheable,
room-temperature ME material that is isotropic in the absence of
any field. It exhibits a ME response which results from the
orientation of the constituent particles. The motion of the
magnetoelectric particles itself is now an integral part of the
material's response, and this may be exploited in applications.
Should the particles be optically anisotropic, as in this
particular case, then ME properties can be studied optically,
which should prove particularly useful in the characterization of
a whole host of potential new magnetoelectric particles, including
colloidal and nano-sized particles. Complex electromagnetic,
rheological, and dynamical effects are expected in this new class
of magnetoelectrics. Finally, the combination of electric and
magnetic moments in a single particle introduces a new handle in
the manipulation of micro- and nanoparticles and is thus expected
to be a generally useful property that should find application in
other fields. \\
Acknowledgements: We thank Winfield Hill for help with the
electronics, David Phillips for the loan of the fluxgate and the
Rowland Institute for financial support. $\ddagger$Permanent
address: Xerox PARC, Palo Alto CA, USA.


\begin{thebibliography}{22}
\expandafter\ifx\csname
natexlab\endcsname\relax\def\natexlab#1{#1}\fi
\expandafter\ifx\csname bibnamefont\endcsname\relax
  \def\bibnamefont#1{#1}\fi
\expandafter\ifx\csname bibfnamefont\endcsname\relax
  \def\bibfnamefont#1{#1}\fi
\expandafter\ifx\csname citenamefont\endcsname\relax
  \def\citenamefont#1{#1}\fi
\expandafter\ifx\csname url\endcsname\relax
  \def\url#1{\texttt{#1}}\fi
\expandafter\ifx\csname
urlprefix\endcsname\relax\def\urlprefix{URL }\fi
\providecommand{\bibinfo}[2]{#2}
\providecommand{\eprint}[2][]{\url{#2}}

\bibitem[{\citenamefont{Fiebig}(2005)}]{fiebig2005}
\bibinfo{author}{\bibfnamefont{M.}~\bibnamefont{Fiebig}}, \bibinfo{journal}{J.
  Phys. D-Appl. Phys.} \textbf{\bibinfo{volume}{38}}, \bibinfo{pages}{R123}
  (\bibinfo{year}{2005}).

\bibitem[{\citenamefont{Eerenstein et~al.}(2006)\citenamefont{Eerenstein,
  Mathur, and Scott}}]{eerenstein2006}
\bibinfo{author}{\bibfnamefont{W.}~\bibnamefont{Eerenstein}},
  \bibinfo{author}{\bibfnamefont{N.~D.} \bibnamefont{Mathur}},
  \bibnamefont{and} \bibinfo{author}{\bibfnamefont{J.~F.} \bibnamefont{Scott}},
  \bibinfo{journal}{Nature} \textbf{\bibinfo{volume}{442}},
  \bibinfo{pages}{759} (\bibinfo{year}{2006}).

\bibitem[{\citenamefont{Nan}(1994)}]{nan1994}
\bibinfo{author}{\bibfnamefont{C.~W.} \bibnamefont{Nan}},
  \bibinfo{journal}{Phys. Rev. B} \textbf{\bibinfo{volume}{50}},
  \bibinfo{pages}{6082} (\bibinfo{year}{1994}).

\bibitem[{\citenamefont{van Vleck}(1985)}]{Vleck1985}
\bibinfo{author}{\bibfnamefont{J.~H.} \bibnamefont{van Vleck}},
  \emph{\bibinfo{title}{{T}he theory of electric and magnetic
  susceptibilities}} (\bibinfo{publisher}{Oxford Univ. Press},
  \bibinfo{year}{1985}).

\bibitem[{tn1()}]{tn1}
\bibinfo{note}{See online supplementary information.}

\bibitem[{\citenamefont{Tellegen}(1948)}]{tellegen1948}
\bibinfo{author}{\bibfnamefont{B.~D.~H.} \bibnamefont{Tellegen}},
  \bibinfo{journal}{Philips Res. Rep.} \textbf{\bibinfo{volume}{3}},
  \bibinfo{pages}{81} (\bibinfo{year}{1948}).

\bibitem[{\citenamefont{O'Dell}(1970)}]{odell}
\bibinfo{author}{\bibfnamefont{T.~H.} \bibnamefont{O'Dell}},
  \emph{\bibinfo{title}{The electrodynamics of magneto-electric media}},
  vol.~\bibinfo{volume}{11} of \emph{\bibinfo{series}{series monographs on
  selected topics in solid state physics}} (\bibinfo{publisher}{North-Holland},
  \bibinfo{year}{1970}).

\bibitem[{\citenamefont{Raab and Sihvola}(1997)}]{raab1997}
\bibinfo{author}{\bibfnamefont{R.~E.} \bibnamefont{Raab}} \bibnamefont{and}
  \bibinfo{author}{\bibfnamefont{A.~H.} \bibnamefont{Sihvola}},
  \bibinfo{journal}{J. Phys. A-Math.} \textbf{\bibinfo{volume}{30}},
  \bibinfo{pages}{1335} (\bibinfo{year}{1997}).

\bibitem[{\citenamefont{Saha et~al.}(2002)\citenamefont{Saha, Kamenetskii, and
  Awai}}]{saha2002}
\bibinfo{author}{\bibfnamefont{A.~K.} \bibnamefont{Saha}},
  \bibinfo{author}{\bibfnamefont{E.~O.} \bibnamefont{Kamenetskii}},
  \bibnamefont{and} \bibinfo{author}{\bibfnamefont{I.}~\bibnamefont{Awai}},
  \bibinfo{journal}{J. Phys. D-Appl. Phys.} \textbf{\bibinfo{volume}{35}},
  \bibinfo{pages}{2484} (\bibinfo{year}{2002}).

\bibitem[{\citenamefont{Tretyakov et~al.}(2003)\citenamefont{Tretyakov,
  Maslovski, Nefedov, Vitanen, Belov, and Sanmartin}}]{tretyakov2003}
\bibinfo{author}{\bibfnamefont{S.~A.} \bibnamefont{Tretyakov}},
  \bibinfo{author}{\bibfnamefont{S.~I.} \bibnamefont{Maslovski}},
  \bibinfo{author}{\bibfnamefont{I.~S.} \bibnamefont{Nefedov}},
  \bibinfo{author}{\bibfnamefont{A.~J.} \bibnamefont{Vitanen}},
  \bibinfo{author}{\bibfnamefont{P.~A.} \bibnamefont{Belov}}, \bibnamefont{and}
  \bibinfo{author}{\bibfnamefont{A.}~\bibnamefont{Sanmartin}},
  \bibinfo{journal}{Electrom.} \textbf{\bibinfo{volume}{23}},
  \bibinfo{pages}{665} (\bibinfo{year}{2003}).

\bibitem[{\citenamefont{Zhai et~al.}(2006)\citenamefont{Zhai, Li, Dong,
  Viehland, and Bichurin}}]{zhai2006}
\bibinfo{author}{\bibfnamefont{J.~Y.} \bibnamefont{Zhai}},
  \bibinfo{author}{\bibfnamefont{J.~F.} \bibnamefont{Li}},
  \bibinfo{author}{\bibfnamefont{S.~X.} \bibnamefont{Dong}},
  \bibinfo{author}{\bibfnamefont{D.}~\bibnamefont{Viehland}}, \bibnamefont{and}
  \bibinfo{author}{\bibfnamefont{M.~I.} \bibnamefont{Bichurin}},
  \bibinfo{journal}{J. Appl. Phys.} \textbf{\bibinfo{volume}{100}},
  \bibinfo{pages}{124509} (\bibinfo{year}{2006}).

\bibitem[{\citenamefont{Lindell et~al.}(1994)\citenamefont{Lindell, Sihvola,
  Tretyakov, and Viitanen}}]{lindell1994}
\bibinfo{author}{\bibfnamefont{I.}~\bibnamefont{Lindell}},
  \bibinfo{author}{\bibfnamefont{A.}~\bibnamefont{Sihvola}},
  \bibinfo{author}{\bibfnamefont{S.}~\bibnamefont{Tretyakov}},
  \bibnamefont{and} \bibinfo{author}{\bibfnamefont{A.}~\bibnamefont{Viitanen}},
  \emph{\bibinfo{title}{{E}lectromagnetic waves in chiral and bi-isotropic
  media}} (\bibinfo{publisher}{Artech House}, \bibinfo{address}{London},
  \bibinfo{year}{1994}).

\bibitem[{\citenamefont{Takei and Shimizu}(1997)}]{takei1997}
\bibinfo{author}{\bibfnamefont{H.}~\bibnamefont{Takei}} \bibnamefont{and}
  \bibinfo{author}{\bibfnamefont{N.}~\bibnamefont{Shimizu}},
  \bibinfo{journal}{Langmuir} \textbf{\bibinfo{volume}{13}},
  \bibinfo{pages}{1865} (\bibinfo{year}{1997}).

\bibitem[{\citenamefont{Anker and Kopelman}(2003)}]{anker2003}
\bibinfo{author}{\bibfnamefont{J.~N.} \bibnamefont{Anker}} \bibnamefont{and}
  \bibinfo{author}{\bibfnamefont{R.}~\bibnamefont{Kopelman}},
  \bibinfo{journal}{Appl. Phys. Lett.} \textbf{\bibinfo{volume}{82}},
  \bibinfo{pages}{1102} (\bibinfo{year}{2003}).

\bibitem[{\citenamefont{Perro et~al.}(2005)\citenamefont{Perro, Reculusa,
  Ravaine, Bourgeat-Lami, and Duguet}}]{perro2005}
\bibinfo{author}{\bibfnamefont{A.}~\bibnamefont{Perro}},
  \bibinfo{author}{\bibfnamefont{S.}~\bibnamefont{Reculusa}},
  \bibinfo{author}{\bibfnamefont{S.}~\bibnamefont{Ravaine}},
  \bibinfo{author}{\bibfnamefont{E.~B.} \bibnamefont{Bourgeat-Lami}},
  \bibnamefont{and} \bibinfo{author}{\bibfnamefont{E.}~\bibnamefont{Duguet}},
  \bibinfo{journal}{J. Mat. Chem.} \textbf{\bibinfo{volume}{15}},
  \bibinfo{pages}{3745} (\bibinfo{year}{2005}).

\bibitem[{\citenamefont{Crowley et~al.}(2002)\citenamefont{Crowley, Sheridon,
  and Romano}}]{crowley2002}
\bibinfo{author}{\bibfnamefont{J.~M.} \bibnamefont{Crowley}},
  \bibinfo{author}{\bibfnamefont{N.~K.} \bibnamefont{Sheridon}},
  \bibnamefont{and} \bibinfo{author}{\bibfnamefont{L.}~\bibnamefont{Romano}},
  \bibinfo{journal}{J. Electrostat.} \textbf{\bibinfo{volume}{55}},
  \bibinfo{pages}{247} (\bibinfo{year}{2002}).

\bibitem[{\citenamefont{Nisisako et~al.}(2006)\citenamefont{Nisisako, Torii,
  Takahashi, and Takizawa}}]{nisisako2006}
\bibinfo{author}{\bibfnamefont{T.}~\bibnamefont{Nisisako}},
  \bibinfo{author}{\bibfnamefont{T.}~\bibnamefont{Torii}},
  \bibinfo{author}{\bibfnamefont{T.}~\bibnamefont{Takahashi}},
  \bibnamefont{and} \bibinfo{author}{\bibfnamefont{Y.}~\bibnamefont{Takizawa}},
  \bibinfo{journal}{Adv. Mat.} \textbf{\bibinfo{volume}{18}},
  \bibinfo{pages}{1152} (\bibinfo{year}{2006}).

\bibitem[{\citenamefont{Cayre et~al.}(2003)\citenamefont{Cayre, Paunov, and
  Velev}}]{cayre2003}
\bibinfo{author}{\bibfnamefont{O.}~\bibnamefont{Cayre}},
  \bibinfo{author}{\bibfnamefont{V.~N.} \bibnamefont{Paunov}},
  \bibnamefont{and} \bibinfo{author}{\bibfnamefont{O.~D.} \bibnamefont{Velev}},
  \bibinfo{journal}{Chem. Comm.} pp. \bibinfo{pages}{2296--2297}
  (\bibinfo{year}{2003}).

\bibitem[{\citenamefont{Correa-Duarte et~al.}(2005)\citenamefont{Correa-Duarte,
  Salgueirino-Maceira, Rodriguez-Gonzalez, Liz-Marzan, Kosiorek, Kandulski, and
  Giersig}}]{correa-duarte2005}
\bibinfo{author}{\bibfnamefont{M.~A.} \bibnamefont{Correa-Duarte}},
  \bibinfo{author}{\bibfnamefont{V.}~\bibnamefont{Salgueirino-Maceira}},
  \bibinfo{author}{\bibfnamefont{B.}~\bibnamefont{Rodriguez-Gonzalez}},
  \bibinfo{author}{\bibfnamefont{L.~M.} \bibnamefont{Liz-Marzan}},
  \bibinfo{author}{\bibfnamefont{A.}~\bibnamefont{Kosiorek}},
  \bibinfo{author}{\bibfnamefont{W.}~\bibnamefont{Kandulski}},
  \bibnamefont{and} \bibinfo{author}{\bibfnamefont{M.}~\bibnamefont{Giersig}},
  \bibinfo{journal}{Adv. Mat.} \textbf{\bibinfo{volume}{17}},
  \bibinfo{pages}{2014} (\bibinfo{year}{2005}).

\bibitem[{\citenamefont{Sheridon}(2005)}]{sheridon2005}
\bibinfo{author}{\bibfnamefont{N.~K.} \bibnamefont{Sheridon}}, in
  \emph{\bibinfo{booktitle}{Flexible flat panel displays}}, edited by
  \bibinfo{editor}{\bibfnamefont{G.~P.} \bibnamefont{Crawford}}
  (\bibinfo{publisher}{Wiley}, \bibinfo{year}{2005}).

\bibitem[{\citenamefont{Ryu et~al.}(2002)\citenamefont{Ryu, Priya, Uchino, and
  Kim}}]{ryu2002}
\bibinfo{author}{\bibfnamefont{J.}~\bibnamefont{Ryu}},
  \bibinfo{author}{\bibfnamefont{S.}~\bibnamefont{Priya}},
  \bibinfo{author}{\bibfnamefont{K.}~\bibnamefont{Uchino}}, \bibnamefont{and}
  \bibinfo{author}{\bibfnamefont{H.~E.} \bibnamefont{Kim}},
  \bibinfo{journal}{J. Electroceram.} \textbf{\bibinfo{volume}{8}},
  \bibinfo{pages}{107} (\bibinfo{year}{2002}).

\bibitem[{\citenamefont{Lakhtakia and Weiglhofer}(1994)}]{lakhtakia1994b}
\bibinfo{author}{\bibfnamefont{A.}~\bibnamefont{Lakhtakia}} \bibnamefont{and}
  \bibinfo{author}{\bibfnamefont{W.~S.} \bibnamefont{Weiglhofer}},
  \bibinfo{journal}{Phys. Rev. E} \textbf{\bibinfo{volume}{50}},
  \bibinfo{pages}{5017} (\bibinfo{year}{1994}).

\bibitem[{\citenamefont{Hehl and Obukhov}(2005)}]{hehl2005}
\bibinfo{author}{\bibfnamefont{F.~W.} \bibnamefont{Hehl}} \bibnamefont{and}
  \bibinfo{author}{\bibfnamefont{Y.~N.} \bibnamefont{Obukhov}},
  \bibinfo{journal}{Phys. Lett. A} \textbf{\bibinfo{volume}{334}},
  \bibinfo{pages}{249} (\bibinfo{year}{2005}).

\end{thebibliography}
\end{document}